\documentstyle[aas2pp4]{article}   
\twocolumn

\slugcomment{submitted for publication to: \quad {\bf Rev. Mex. Astron. Astrofis.} }

\received{ 1998 December 15}
\accepted{ 1999 July 1}

\lefthead{Garc\'{\i}a-Barreto et al.}
\righthead{Optical Spectroscopy}
\begin{document} 
\input epsf
%===\defet al., {{\it et~al.\ }}
%===\def\eg{{\it e.~g.\ }}
%===\def\ie{{\it i.~e.,\ }}
 
\title{Optical Spectroscopy of the Central Regions of Bright Barred Spiral Galaxies}

\author{J.A. Garc\'{\i}a-Barreto\altaffilmark{1}, H. Aceves\altaffilmark{2}, O. Kuhn\altaffilmark{3}, G. Canalizo\altaffilmark{4}, R. Carrillo\altaffilmark{1} \and J. Franco\altaffilmark{1}}
\altaffiltext{1}{Instituto de Astronom\'{\i}a, Universidad Nacional Aut\'onoma de M\'exico, Apartado Postal 70-264, M\'exico D.F. 04510 M\'exico.
\\
tony,rene,pepe@astroscu.unam.mx}
\altaffiltext{2}{Instituto de Astrof\'{\i}sica de Andaluc\'{\i}a. Apdo. Postal 3004, Granada 18080. Espa\~na. aceves@iaa.es}
\altaffiltext{3}{Observatorio Astron\'omico Nacional, Instituto de Astronom\'{\i}a, Universidad Nacional Aut\'onoma de M\'exico, Apartado Postal 877, Ensenada, Baja California C.P. 22100, M\'exico}
\altaffiltext{4}{Institute for Astronomy, University of Hawaii, 2680 Woodlawn Drive, Honolulu, HI 96822}

\begin{abstract}

{\it Se presentan espectros \'opticos en la banda roja de 18 
galaxias espirales con barra. El estudio se enfoca a la determinaci\'on de la 
cinem\'atica local y condiciones del gas ionizado en el n\'ucleo compacto 
(dentro de un diametro de $5''$) y en las regiones circunnucleares (dentro 
de un diametro de $20''$). S\'olo 8 galaxias presentan emision brillante al 
Este y Oeste del n\'ucleo compacto. En otras 10, las l\'{\i}neas de emisi\'on 
son d\'ebiles y s\'olo pudimos obtener un espectro promedio de la emisi\'on 
central. No se detectaron l\'{\i}neas de emisi\'on en las otras 8 galaxias. Se 
presenta una estimaci\'on de la masa din\'amica en la region central de cada 
galaxia en base a las velocidades observadas en regiones circunnucleares. En 
NGC 4314 y NGC 6951, que presentan emisi\'on de H$\alpha$ distribuida en 
estructuras de anillo alrededor del n\'ucleo compacto, se determinan los 
cocientes [NII]$\lambda6583$/H$\alpha$ y [SII]/H$\alpha$ para ambos lados del 
anillo. La diferencia de velocidades en uno y otro lado se usa como indicativo
de la velocidad de rotaci\'on del gas alrededor del n\'ucleo de la galaxia. Las 
velocidades encontradas en las regiones circunnucleares en NGC 4314 y NGC 6951 
explican naturalmente la discrepancia que existe en la literatura sobre las 
velocidades de recesi\'on reportadas. Se encuentra que los cocientes 
[NII]$\lambda6583$/H$\alpha$ y [SII]/H$\alpha$ son diferentes en cada lado del anillo (un factor de $\sim2$ mayor en el lado oeste) y se infiere que las 
condiciones f\'{\i}sicas son diferentes. El cociente 
[NII]$\lambda6583$/H$\alpha$ en el n\'ucleo de NGC 6951 es un factor de 2 mayor 
comparado con el valor de la region oeste. Las densidades electr\'onicas se han 
estimado de los cocientes de l\'{\i}neas de azufre [SII].}

Optical red spectra of a set of 18 bright barred spiral galaxies 
are presented. The study is aimed at determining the local kinematics, and 
physical conditions of ionized gas in the compact nucleus (inside a diameter of 
$5''$) and in the circumnuclear regions (inside a diameter of $20''$). Only 8 
galaxies showed bright emission from their east and west side of the nucleus. 
The spectrum of each region was analized separately. In  other 10 galaxies the 
line emission was so weak that we were only able to obtain an average spectrum 
of the central emission. No emission was detected in the remaining 8 galaxies.
An estimate of the dynamical mass is presented based on the observed velocities 
in the circumnuclear regions. In NGC 4314 and NGC 6951, that show H$\alpha$ emission distributed in circumnuclear ring structures, we determine the 
[NII]/H$\alpha$ and [SII]/H$\alpha$ ratios for the eastern and western regions of the rings. The velocity difference for the two sides is used to derive the
rotation velocity of the gas around the compact nucleus. The ratio, 
[NII]$\lambda6583$/H$\alpha$, is a factor of 2 larger in the compact nucleus of  NGC 6951 than in its western side. The electron gas densities have been 
estimated from the [SII] lines ratio. 

\end{abstract}

\section {Introduction}

The study of emission lines in spiral galaxies has  proven to be very useful in 
determining the distribution of ionized gas, and the physical conditions at the
central regions (Burbidge \& Burbidge 1960a, 1960b, 1962, 1964, 1965;  Burbidge, Burbidge \& Pendergast 1960; Keel 1983a,b;
Kennicutt \& Kent 1983; Heckman, Balick \& Crane 1980; Stauffer 1982a,b; 
Kennicutt 1992a,b; Appenzeller \& \"{O}streicher 1988; Filippenko \& Sargent 1985; Ho, Filippenko \& Sargent 1995, 1997a,b; Vaceli et al. 1997). 

Optical observations of barred spiral galaxies usually show the presence of 
symmetric structures around their center (presumably in the plane of the disk) 
which are referred to as {\it rings} (e.g. \cite{but86,but91,but95}). These 
symmetric structures are observed at different radii in their host galaxy; some 
are seen as extended structures at $\approx 10-15$ kpc from the center, but 
some others are located about  $4-5$ kpc, and there are circumnuclear rings at 
$300-900$ pc from the compact nucleus (e.g. \cite{ars88,pog89a},1989b;  
\cite{gar91a},b,1996; \cite{bar95,gen95}). A dynamical explanation for the 
origin of the structures has been sought in terms of non-symmetric 
perturbations to the `normal' gravitational potential of spiral galaxies 
(\cite{bin87}). The standard model to explain the formation of circumnuclear 
structures requieres knowledge of the angular velocity of the gas and stars 
($\Omega_g$), the epiclyclic frequencies ($\kappa$), and an estimate of the 
angular velocity of the bar ($\Omega_b$). In this model, it is generally 
assumed that the gravitational potential of the bar and $\Omega_b$ are 
independent of time (\cite{bin87}). Some of these galaxies have nearby 
neighbors and many numerical models  have been carried out to examine the 
possible role of tidal interactions in the formation of bars and  symmetric 
structures (e.g. \cite{nog88,fri93}). 

Previous spectra of Shapley Ames Galaxies (\cite{fil85,ho95,ho97a,ho97b} represents the emission of the innermost $2''\times7''$ or r $\leq$ 200 pc. Here we present new optical spectroscopic observations of the central regions of 18 barred galaxies using a long slit. Our main purpose is to 
provide local velocities and line ratios from the emission lines of [NII], 
H$\alpha$, and [SII], for both the nucleus and the circumnuclear regions. We 
have tried to obtain spectra of the compact nucleus and from the immediate 
surroundings separately (in the east and west directions since the slit was 
oriented at PA$\sim90^{\circ}$ E of N) by adding only the corresponding pixels 
in the slit. A clear example of this is in NGC 6951, where the spatial 
distribution of H$\alpha$ shows the compact nucleus in addition to a 
circumnuclear ring. NGC 4314 also presents the circumnuclear ring in H$\alpha$, 
but no emission from the compact nucleus is detected in this case. In other 
six galaxies the spatial distribution of the H$\alpha$ emission is more 
diffuse. In 10 galaxies we were only able to obtain a spectrum of the central 
emission: compact nucleus and weak (if present at all) extranuclear emission.

The present work is divided as follows. In section 2, the selection criteria 
for the set of galaxies observed are established, and the observations are 
described. In section 3, our results are presented and discussed. In section 4,
a brief summary is presented.

% Table 1
%\small
\begin{deluxetable}{lrrrrrcc}
%\footnotesize
\scriptsize
%\begin{flushleft}
\tablecaption{Observed Galaxies. \label{tbl-1}}
\tablewidth{0pt}
\tablehead{
\colhead{Galaxy\tablenotemark{a}} & \colhead{Optical\tablenotemark{b}}   
& \colhead{HI\tablenotemark{c}}   & \colhead{CO\tablenotemark{d}}  
& \colhead{{\it i}\tablenotemark{e}} &\colhead{PA$_{bar}$\tablenotemark{e}}
&\colhead{CiNE\tablenotemark{f}}
&\colhead{CoNE\tablenotemark{f}}
}
\startdata 
3504 & 1535$\pm19$  & 1538      &  1550 & 35 & 145 & N & Y\nl
4123 & 1325$\pm10$  & 1327      &       & 39 & 105 & N & Y\nl
4314 &  883$\pm85$  &  982      &  985  & 30 & 145 & Y & N\nl
%4385 & 2142$\pm7$   & 2140      &  2144 &    & 100 & N & Y\nl
%4435 &  869$\pm100$ &  776      &       &    &  15 & ? & ?\nl
4477 & 1263$\pm75$  & 1263      &       & 26 &  15 & ? & ?\nl
%4507 & 3513$\pm25$  & 3525      &       &    &  50 & N & Y\nl
%4561 & 1407$\pm20$  & 1407      &       &    & 120 & N & Y\nl
%4688 &  981$\pm9$   &  981      &       &    &  35 & N & N\nl
4691 & 1123$\pm13$  & 1119      &  1124 & 32 &  85 & Y? & Y?\nl
5135 & 4157$\pm35$  & 4157      &       & 67 & 125 & Y & Y\nl
%5188 & 2366$\pm43$  & 2326      &       &    &  50 & N & Y\nl
5347 & 2373$\pm12$  & 2386      &       & 37 & 100 & Y & Y\nl
5383 & 2260$\pm3$   & 2264      &       & 40 & 130 & Y? & Y\nl
5430 & 2875$\pm90$  & 2960      & 2981  &    & 145 & N & Y\nl
5534 & 2633$\pm10$  & 2633      &       &    &  80 & N & Y\nl
5597 & 2624$\pm66$  &           &       &    &  55 & Y & Y\nl
5691 & 1876$\pm50$  &           &       & 34 &  90 & Y? & Y\nl
5728 & 2970$\pm40$  & 2780      &       & 65 &  35 & Y & Y?\nl
5757 & 2771$\pm58$  &           &       & 32 & 165 & Y & N\nl
5915 & 2273$\pm14$  & 2274      &       & 42 &  90 & Y? & Y?\nl
6239 &  938$\pm12$  &  931      &       & 67 & 115 & N & Y?\nl
6907 & 3155$\pm80$  & 3139      &       &    &  45 & Y & Y\nl
6951 & 1425$\pm10$  & 1426      &  1464 & 28 &  85 & Y & Y\nl
%7371 & 2685$\pm15$  & 2685      &       &    &  10 & ? & ?\nl
%5273 & 1293$\pm15$  & 1298      &       &    &     & Y & Y?\nl
\enddata
%\end{flushleft}
% Text for table footnotes follows the tabular data and must be inside the
% deluxetable environment.  Note that it is OK to put \ref's in 
% \tablenotetext's.
\tablenotetext{a}{NGC catalog numbers} 
\tablenotetext{b}{Systemic Velocity in km $s^{-1}$ from Revised Shapley Ames Catalog, (Sandage \& Tammann 1987)}
\tablenotetext{c}{Systemic Velocity in km $s^{-1}$ from Huchtmeier \& Richter (1989),except for NGC 4314 which we used the value observed by Garcia-Barreto, Downes \& Huchtmeier 1994}
\tablenotetext{d}{Systemic Velocity in km $s^{-1}$ from Young et al., (1995) for most of the galaxies except for NGC 4314 taken from Garcia-Barreto et al., (1991b)}
\tablenotetext{e}{Inclination of the galaxy's disk with respect to the palne of the sky; position angle of the stellar bar measured East of North taken from Tully (1988) or Huchtmeier \& Richter (1989)}
\tablenotetext{f}{Observed spatial H$\alpha$+[NII] emission : Circumnuclear or extranuclear (innermost 20$''$) (CiNE); compact nucleus 
(innermost 5$''$)(CoNE) from Pogge (1989a,b), Garcia-Barreto et al., (1996)}
\end{deluxetable}

\section{Sample and Observations}

The galaxies studied here are part of an ongoing study of barred galaxies (see 
Garc\'{\i}a-Barreto et al. 1993,1996), and they were chosen using the following criteria: 1) bright barred galaxies in the Shapley Ames Catalog (first or second edition 
\cite{san87}), 2) with declinations within a range accessible from San Pedro 
M\'artir, $-41^{\circ} {\leq} {\delta} {\leq}+70^{\circ}$, and 3) with IRAS 
colors characteristic of star-forming galaxies according to Helou (1986; with 
log(f(12)/f(25)) ${\leq}-0.15$ and log(f(60)/f(100)) ${\leq}-0.1$) or 
equivalently, an IRAS dust temperature T$_d {\geq}25$ K. The characteristics of 
each galaxy such as distance, IRAS fluxes, radio continuum fluxes, dust 
temperature, blue luminosity, HI mass, X-~Ray luminosity and inclination, are
summarized in Garc\'{\i}a-Barreto et al. (1993,1996). The purpose of the 
observations was to determine the physical conditions (velocities and line 
ratios) from the innermost central regions (circumnuclear structure and compact 
nucleus). The slit length and width were selected accordingly. 
%In no moment had we thought of obtaining the velocity information from 
%the full disk of each galaxy in order to construct rotation curves.
Table 1 lists the systemic velocities from optical, HI and CO observations 
taken from the literature, as well as whether a given galaxy shows H$\alpha$ 
emission from either the compact nuclear region (innermost $5''$ in diameter) 
and or circumnuclear regions (innermost $20''$ in diameter) (ie. 
\cite{pog89a,pog89b,gar96}).

\begin{figure}[!t]
\epsfxsize=8cm
\epsffile{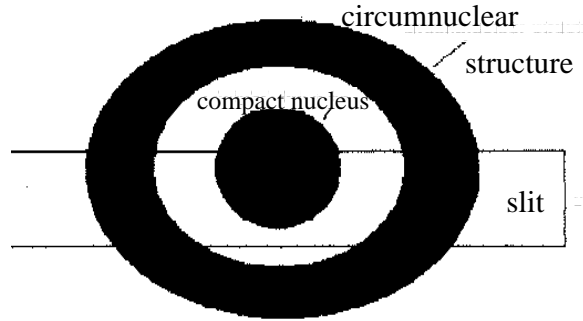}
\caption{ Sketch (not to scale) of the slit position relative to a circumnuclear region in the central region of a given galaxy. The slit in all cases 
positioned at PA 90$^{\circ}$ or east - west. Final length of slit was 59 pixels long or 43$''$, and 10 pixels wide or 7$''$.3. }
\label{fig1}
\end{figure}

% Table 2
\begin{deluxetable}{lccrcrrrrcrcrc}
%\scriptsize
\tiny
\tablecaption{Emission Lines of Galaxies with Circumnuclear Regions. \label{tbl-2}}
\tablewidth{0pt}
\tablehead{
\colhead{Galaxy} & \colhead{Region} & \colhead{Pix\tablenotemark{a}}& \colhead{[NII]\tablenotemark{b}}
& \colhead{[NII]/H$\alpha$} & \colhead{H${\alpha}$\tablenotemark{b}} & \colhead{FWHM\tablenotemark{c}}   & \colhead{H$\alpha$ Flux\tablenotemark{d}}
& \colhead{[NII]\tablenotemark{b}}  & \colhead{[NII]/H$\alpha$}         & \colhead{[SII]\tablenotemark{b}}  & \colhead{[SII]/H$\alpha$}         & \colhead{[SII]\tablenotemark{b}}  & \colhead{[SII]/H$\alpha$}
}
\startdata 
3504 & E & 6 & 1385  & 0.17  & 1383  &  215 & 12.0 & 1391 & 0.53 & 1403 & 0.13 & 1413 & 0.12\nl
     & W & 3 & 1475  & 0.17 & 1470  &  237 &  9.7 & 1476 & 0.57 & 1498 & 0.13 & 1489 & 0.13 \nl
     & N & 6 & 1430  & 0.17 & 1422  &  253 & 11.0 & 1434 & 0.60 & 1440 & 0.12 & 1445 & 0.13\nl
4314 & E & 12 &  935  & 0.13 &  868  &  237 &   1.6 &  853 & 0.45 &  903  & 0.09 &  913 & 0.18\nl
     & W & 11 &  997  & 0.21 & 983  &  224 &  1.0 &  945 & 0.85 & 891 & 0.18 & 938 & 0.25 \nl
4691 & E & 9 & 1017  & $\leq0.01$ & 1050  &  151 &   2.4 & 1062 & 0.38 & 1025 & 0.08 & 1066 & 0.07\nl
     & W & 6 & 1038  & 0.07 & 1055  &  210 &  4.8 & 1059 & 0.41 & 1039 & 0.10 & 1073 & 0.09 \nl
     & W\tablenotemark{e} & 11 &     &      &  635 & 1475& 38.9 &     &     &      &      &  & \nl
     & N & 6 & 1068  & 0.11 & 1068  &  207 & 16.4 & 1081 & 0.42 & 1091 & 0.09 & 1089 & 0.09\nl
5135 & E & 5 & 3996  & 0.15 & 3965  &  279 &   1.1 & 3961 & 0.77 & 3783  & 0.21 &  3954 & 0.16\nl
     & W & 4 & 4021  & 0.29 & 3988  &  247 &  7.1 & 3896 & 0.89 & 3995 & 0.17 & 4024 & 0.19 \nl
     & N & 4 & 3970  & 0.25 & 3966  &  334 & 19.9 & 3954 & 0.76 & 3962 & 0.15 & 3991 & 0.13\nl
5383 & E & 14 & 2589  & 0.10 & 2453  &  224 &   2.3 & 2403 & 0.13 & 2433 & 0.14   &  2829 & 0.10\nl
     & W & 13 & 2240  & 0.06 & 2257  &  219 &  5.4  & 2265 & 0.25 & 2271 
& 0.10 & 2202  & 0.08\nl
     & N & 5  &       &      & 2376  &   56 & 0.8   & 2336 & 0.43 &     
&      &       &\nl
5534 & E & 13 & 2487  & 0.14 & 2477  &  233 &  11.1 & 2483 & 0.45 & 2497 & 0.15 & 2504 & 0.14\nl
     & W & 14 & 2440  & 0.05 & 2502  &  173 &  0.9 & 2510 & 0.27 & 2531 & 0.11 & 2651 & 0.14\nl
5915 & E & 18 & 2192  & 0.09 & 2207  &  196 &   1.5 & 2216 & 0.26 & 2151 & 0.07 &  2348 & 0.09\nl
     & W & 12 & 2094  & 0.16 & 2061  &  201 &  1.6 & 2091 & 0.41 & 2061 & 0.18 & 2097 & 0.14\nl
     & N & 8 & 2148  & 0.18 & 2133  &  259 & 2.9 & 2147 & 0.49 & 2127 & 0.21 & 2185 & 0.14\nl
6951 & E & 6 & 1517  & 0.15 & 1491  &  247 &   1.1 & 1485 & 0.48 & 1652 & 0.09 &  1607& 0.10\nl
     & W & 4 & 1399  & 0.28 & 1348  &  251 &  1.1 & 1371 & 0.93 & 1215 & 0.08 & 1228 & 0.09\nl
     & N & 7 & 1389  & 0.85  & 1428  &  285 & 0.8 & 1397 & 1.99 & 1423 & 0.32  & 1398 & 0.36\nl
\enddata
% Text for table footnotes follows the tabular data and must be inside the
% deluxetable environment.  Note that it is OK to put \ref's in 
% \tablenotetext's.
\tablenotetext{a}{Pixels added in slit for each region to produce the final spectra}
\tablenotetext{b}{Observed heliocentric velocity centroids in km s$^{-1}$. The velocity centroid uncertainties amount to about 25 km s$^{-1}$ as a result of 
wavelength calibration and gaussian fitting. Columns 4 and 5 correspond to  [NII]$~{\lambda}6548.1$ \AA, columns 9 and 10 to [NII]$~{\lambda}6583.4$ \AA, columns 11 and 12 to [SII]$~{\lambda}6716.4$ \AA, and columns 13 and 14 to [SII]$~{\lambda}6730.8$ \AA }
\tablenotetext{c}{H$\alpha$ full width at half maximum, in km s$^{-1}$,  after deconvolution with an instrumental response (FWHM$_{inst}{\sim}3.23\pm0.1$ \AA 
= 145$\pm5$ km s$^{-1}$).}
\tablenotetext{d}{H$\alpha$ flux in units of 10$^{-14}$ ergs s$^{-1}$ cm$^{-2}$}
\tablenotetext{e}{Broadline from a western spectrum 12$''$ away from the spectrum of narrow lines (\cite{gar95})}.

\end{deluxetable}

% Table 3
\begin{deluxetable}{lcrrcrrrrcrcrc}
\tiny
\tablecaption{Emission Lines from Galaxies with only a central spectrum. \label{tbl-3}}
\tablewidth{0pt}
\tablehead{
\colhead{Galaxy} & \colhead{Region} & \colhead{Pix\tablenotemark{a}} & \colhead{[NII]\tablenotemark{b}}  & \colhead{[NII]/H$\alpha$}   & \colhead{H${\alpha}$\tablenotemark{b}} & \colhead{FWHM\tablenotemark{c}}   & \colhead{H$\alpha$ Flux\tablenotemark{d}} 
& \colhead{[NII]\tablenotemark{a}}  & \colhead{[NII]/H$\alpha$}   
& \colhead{[SII]\tablenotemark{a}}  & \colhead{[SII]/H$\alpha$}   
& \colhead{[SII]\tablenotemark{a}}  & \colhead{[SII]/H$\alpha$}
}
\startdata 
4123 & N & 12 & 1211  & 0.15 & 1193  &  251 & 11.3  & 1196 & 0.51 & 1239 & 0.14 & 1215 & 0.12\nl
4477 & N & 13 & 1355  & 0.70  & 1373  & 205 & 0.8 & 1375 & 2.37 & 1410 & 0.39 & 1404 & 0.57\nl 
5347 & N & 10 & 2349  & 0.25 & 2325  &  384 & 2.1 & 2291 & 0.71 & 2349 & 0.34 & 2329 & 0.35\nl
5430 & N & 10 & 3073  & 0.19 & 3055  &  320 & 5.4  & 3051 & 0.54 & 3070 & 0.14 & 3101 & 0.11\nl
5597 & N & 17 & 2491  & 0.15 & 2452  &  269 & 19.0 & 2467 & 0.44 & 2493 & 0.10 & 2605 & 0.06\nl
5691 & N & 13 &       &      & 1733  &  164 &  0.6 &    &     & 1745 & 0.19 & 1831 
& 0.25\nl
5728 & N & 14 & 2852  & 0.45 & 2870  &  507 & 6.2 & 2850 & 1.39 & 2857 & 0.40 & 2865 & 0.32\nl
5757 & N & 5 & 2517  & 0.14 & 2505  &  283 & 15.9 & 2516 & 0.45 & 2530 & 0.11 & 2533 & 0.11\nl
6239 & N & 20 &      &      & 1060  &  132 & 4.8  & 1061 & 0.12 & 1061 & 0.10 &     1072 & 0.05\nl
6907 & N & 18 & 2968  & 0.16 & 2940  &  347 & 9.2 & 2947 & 0.45 & 2925 & 0.09 & 
3002 & 0.07\nl
\enddata
% Text for table footnotes follows the tabular data and must be inside the
% deluxetable environment.  Note that it is OK to put \ref's in 
% \tablenotetext's.
\tablenotetext{a}{Pixels added in slit for central region to produce the final spectrum}
\tablenotetext{b}{Observed heliocentric velocity centroids in km s$^{-1}$. The velocity centroid uncertainties amount to about 25 km s$^{-1}$ as a result of 
wavelength calibration and gaussian fitting. Columns 4 and 5 correspond to  [NII]$~{\lambda}6548.1$ \AA, columns 9 and 10 to [NII]$~{\lambda}6583.4$ \AA, columns 11 and 12 to [SII]$~{\lambda}6716.4$ \AA, and columns 13 and 14 to [SII]$~{\lambda}6730.8$ \AA }
\tablenotetext{c}{H$\alpha$ full width at half maximum, in km s$^{-1}$, after deconvolution with an instrumental response 
(FWHM$_{inst}{\sim}3.23\pm0.1$ \AA = 145$\pm5$ km s$^{-1}$).}
\tablenotetext{d}{H$\alpha$ flux in units of 10$^{-14}$ ergs s$^{-1}$ cm$^{-2}$}
\end{deluxetable}

The optical observations were performed at the Observatorio Astron\'omico 
Nacional in San Pedro M\'ar\-tir, Baja California, M\'exico, on 1993, May 28, 
29, 30, 31 and June 1, using the 2.12m telescope f/7.5 equipped with a 1024 
$\times$ 1024 Thompson CCD detector coupled to a Boller \& Chivens spectrograph 
with 600 grooves per mm at a grating angle of 13$^{\circ}$. The central 
wavelength was chosen in the red part of the spectrum in order to include the 
[N II], H${\alpha}$ and [S II] lines. The detector scale gives  a 1.6 \AA \ 
pixel$^{-1}$ spectral resolution and a spatial resolution of $0''.73$
pixel$^{-1}$ in the spectroscopic mode. The original slit length was 80 pixels 
long (or 58$''$) by 10 pixels (or 7.3$''$) wide positioned at P.A. 
${\sim}90^{\circ}$ (E of N), that is, East -- West. In order to avoid edge 
effects, we discarded in the final analysis altogether the outer 21 pixels from 
the slit length and thus the final slit length was only 59 pixels or 43$''$. 
Only emission from about 19$''$ on each side of a central region of a galaxy 
was observed assuming that the compact nucleus was positioned in the middle of 
the slit covering the innermost 7 pixels.

The slit was positioned on the brightest optical region of each galaxy, at the 
declination corresponding to the compact nucleus. Figure~\ref{fig1} shows a sketch of  the relative position of the slit with respect to a circumnuclear structure in  the innermost central region of a galaxy. Our final analysis showed that in some galaxies the slit was off nucleus by a few arcseconds.

The on source time was 1200s for all the galaxies in the list (except for NGC 
5915 which was observed for only 900s). Each galaxy observation was followed by 
a 1 second exposure of a tungsten lamp. Dome flats taken with the same 
instrumental setup were used to flatten each galaxy spectrum. The data were 
bias-subtracted and flat-fielded using the NOAO IRAF software. The sky was 
subtracted with the IRAF task background using for each galaxy the apertures 
without any line emission (usually on the edges of the slit).

\begin{figure}[!t]
\epsfxsize=8cm
\epsffile{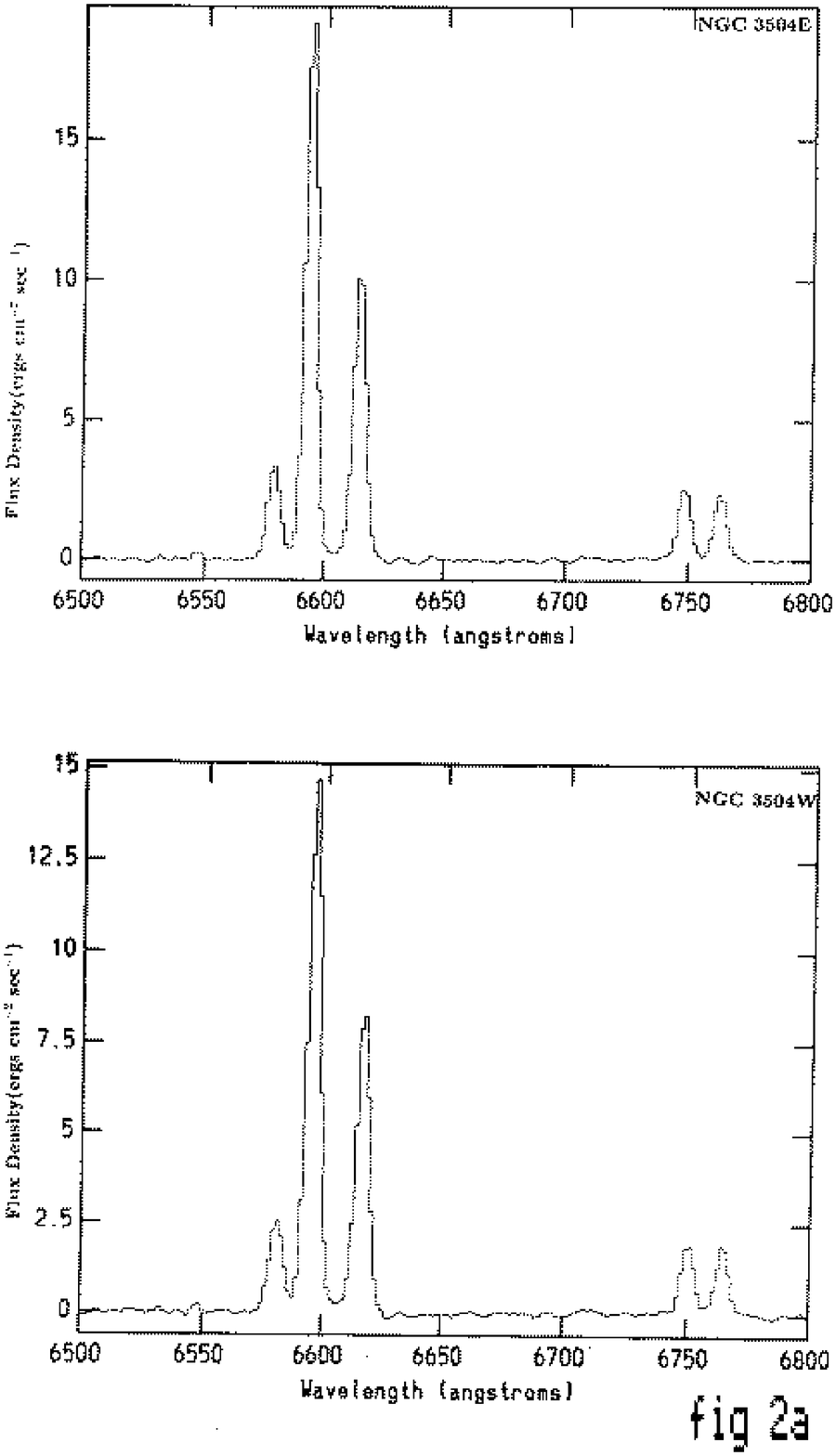}
\caption{   Spectra from the eastern (upper) and western (lower) circumnuclear regions of bright barred spiral galaxies calibrated in wavelength, sky, flux and redshift. The position of the slit was always at the P.A. 90$^{\circ}$. Flux units are $10^{-15}$~ergs s$^{-1}$~cm$^{-2}$~\AA$^{-1}$. Here for NGC 3504. }
\label{fig2a}
\end{figure}

\begin{figure}[!ht]
\epsfxsize=8cm
\epsffile{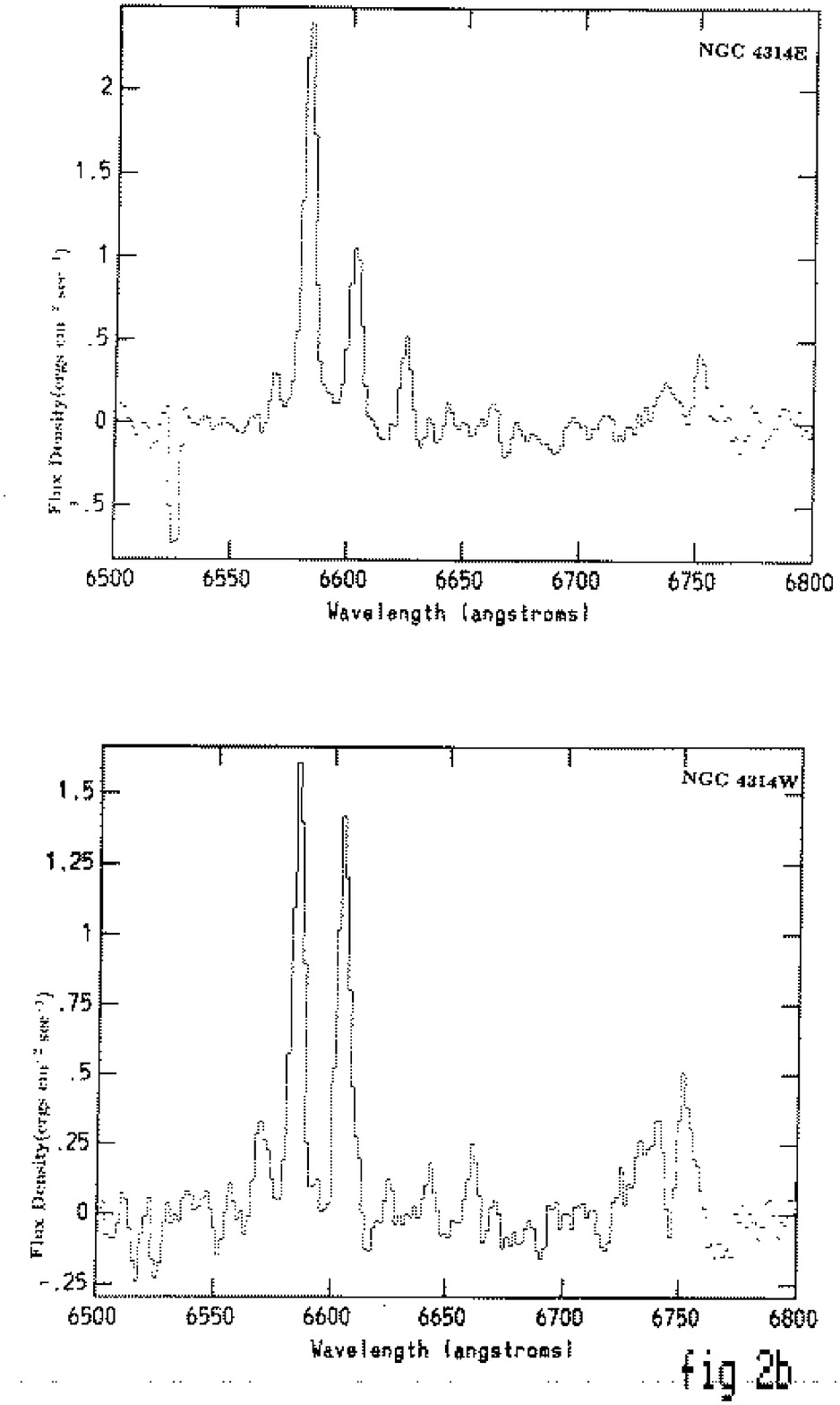}
\caption{As in Fig. 2, but for NGC 4314.}
\label{fig2b}
\end{figure}

\begin{figure}[!ht]
\epsfxsize=8cm
\epsffile{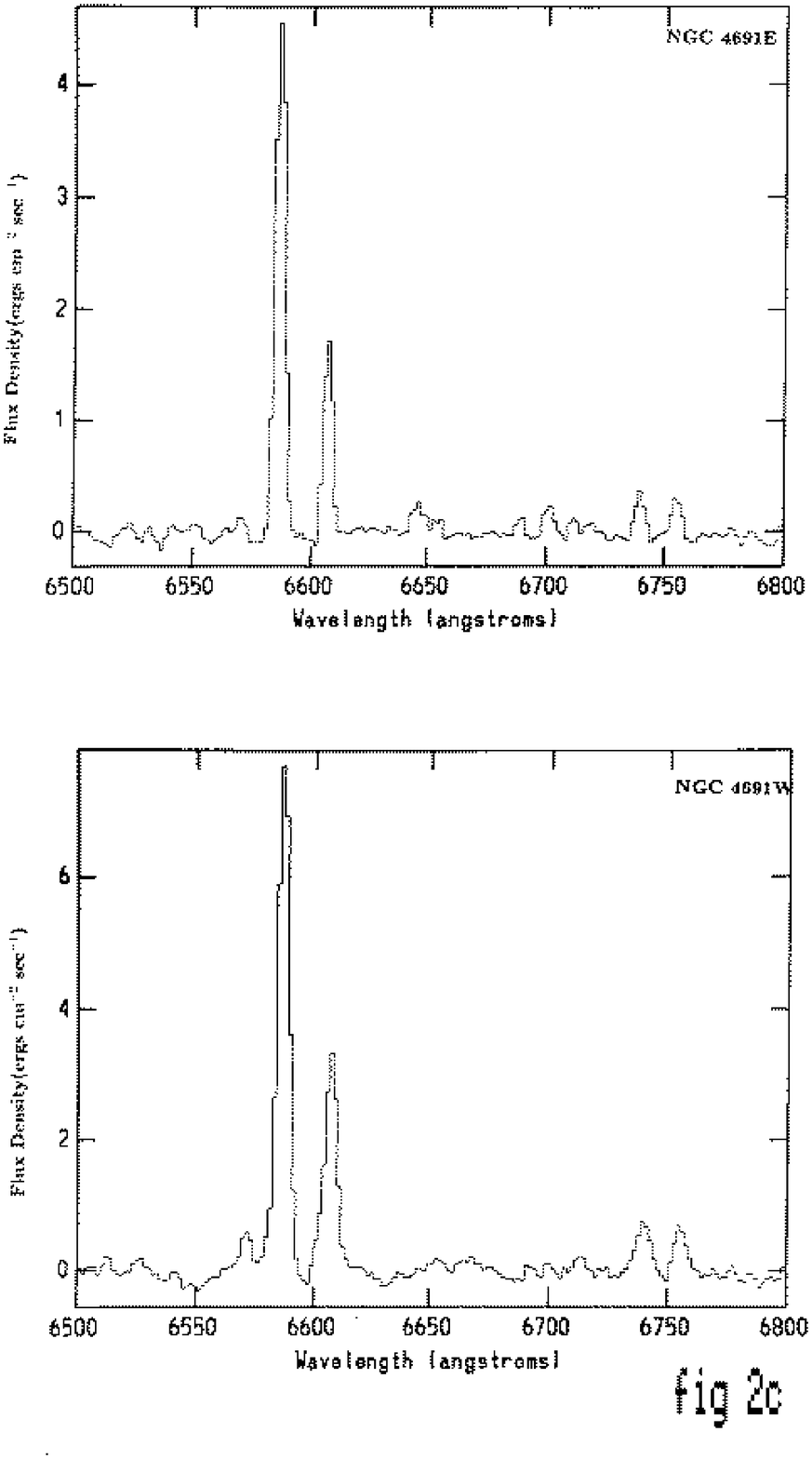}
\caption{As in Fig. 2, but for NGC 4691.}
\label{fig2c}
\end{figure}

\begin{figure}[!ht]
\epsfxsize=8cm
\epsffile{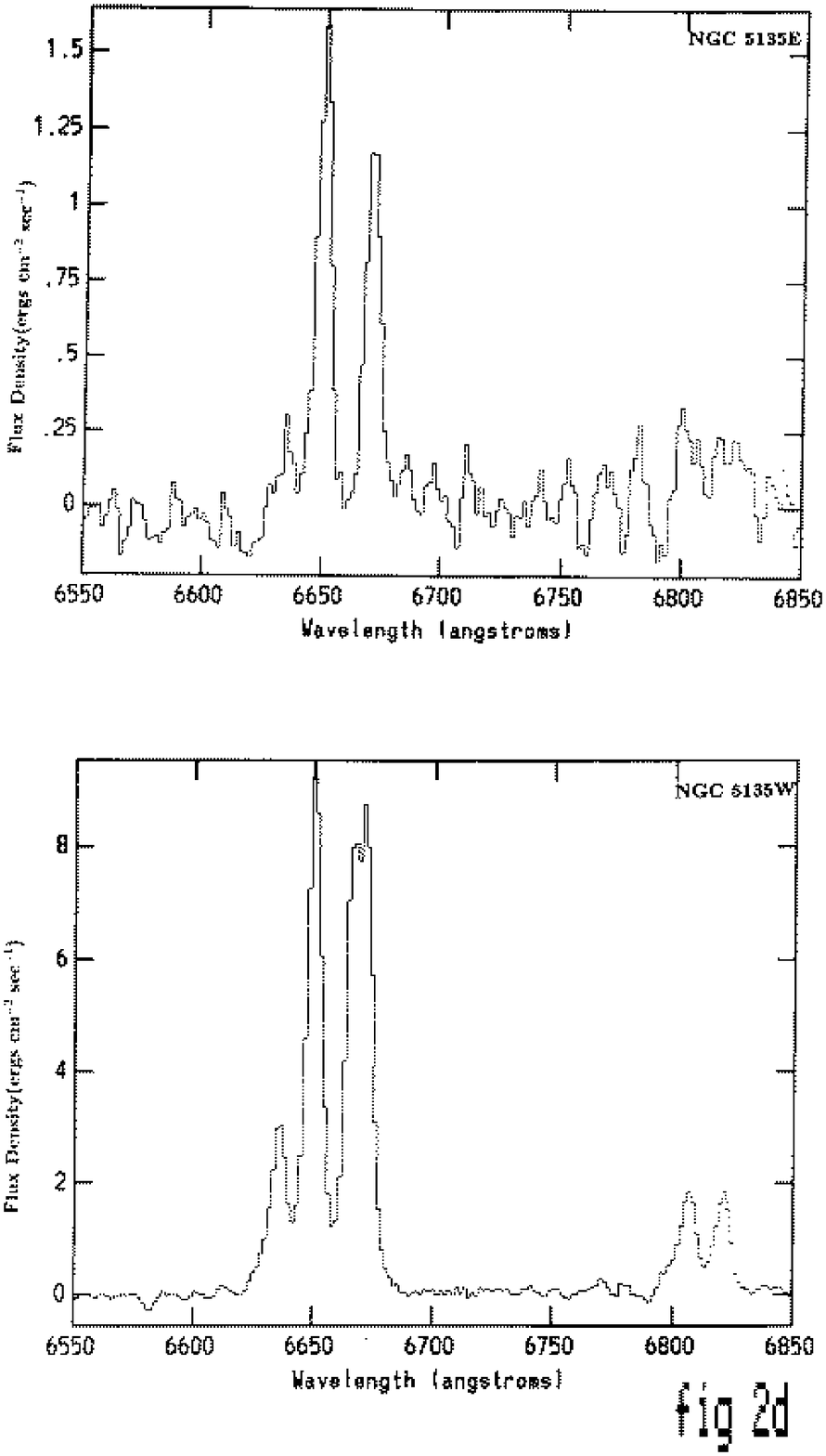}
\caption{As in Fig. 2, but for NGC 5135.}
\label{fig2d}
\end{figure}

\begin{figure}[!ht]
\epsfxsize=8cm
\epsffile{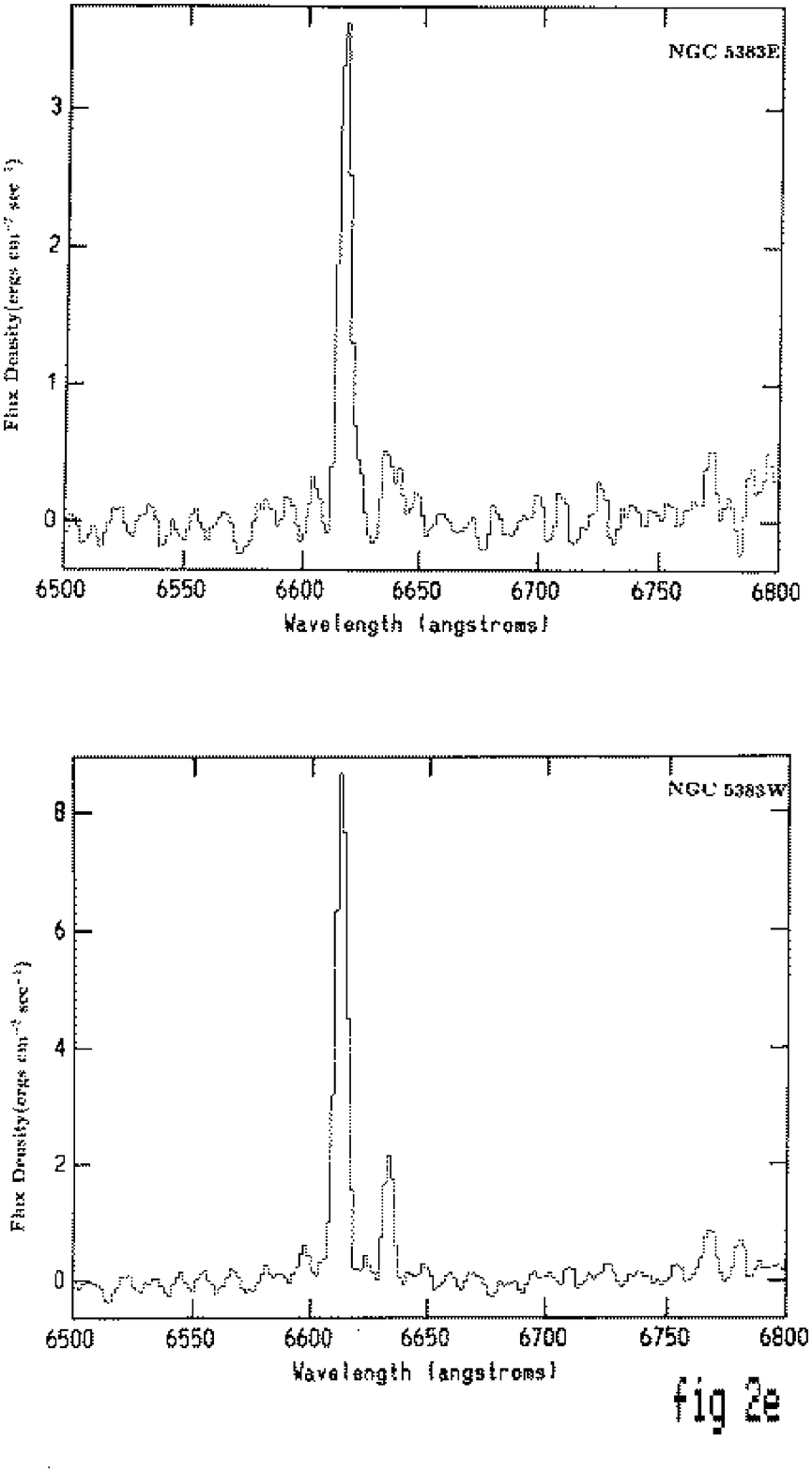}
\caption{As in Fig. 2, but for NGC 5383.}
\label{fig2e}
\end{figure}

\begin{figure}[!ht]
\epsfxsize=8cm
\epsffile{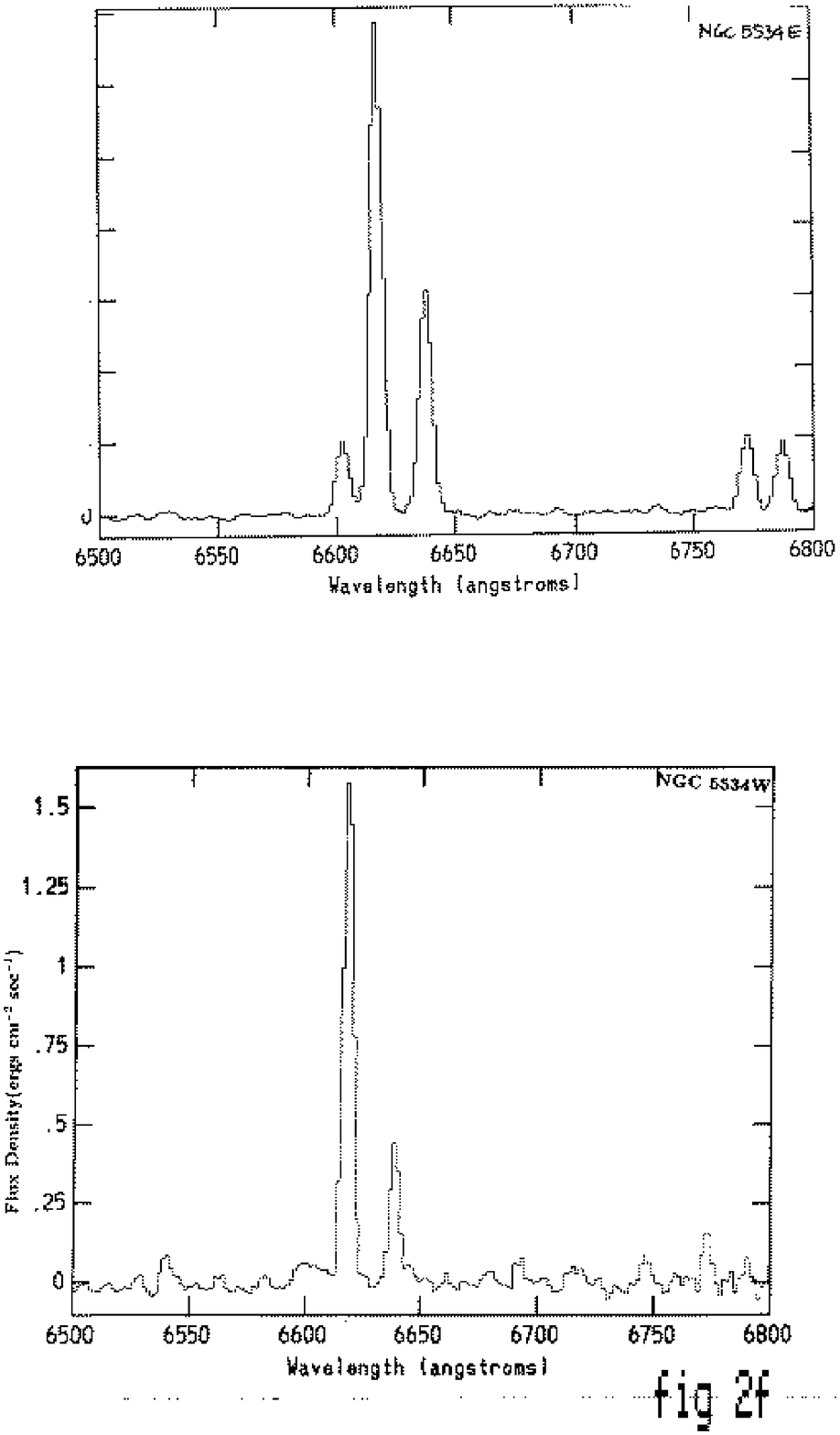}
\caption{As in Fig. 2, but for NGC 5534.}
\label{fig2f}
\end{figure}

\begin{figure}[!ht]
\epsfxsize=8cm
\epsffile{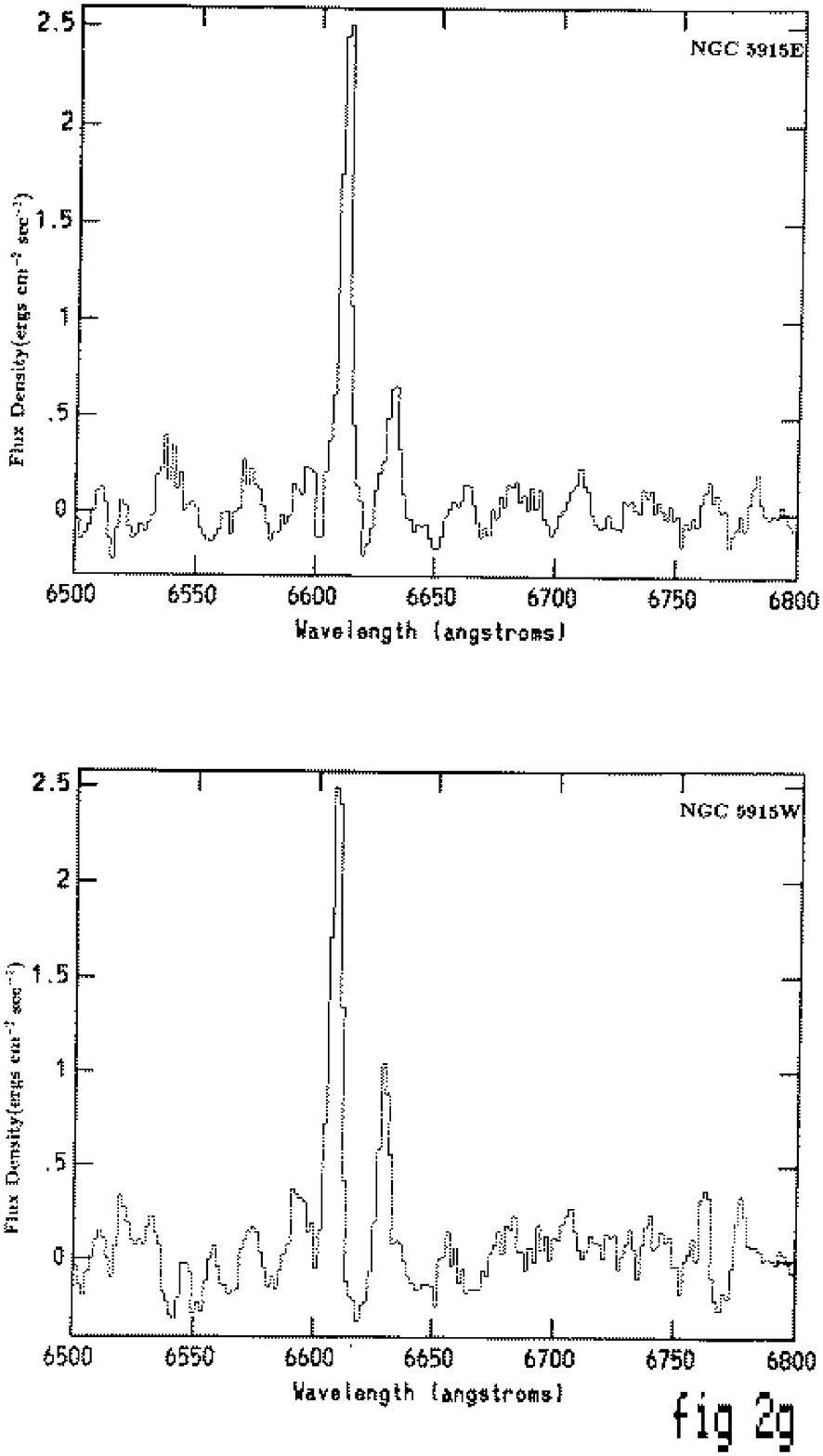}
\caption{As in Fig. 2, but for NGC 5915.}
\label{fig2g}
\end{figure}

\begin{figure}[!ht] 
\epsfxsize=8cm
\epsffile{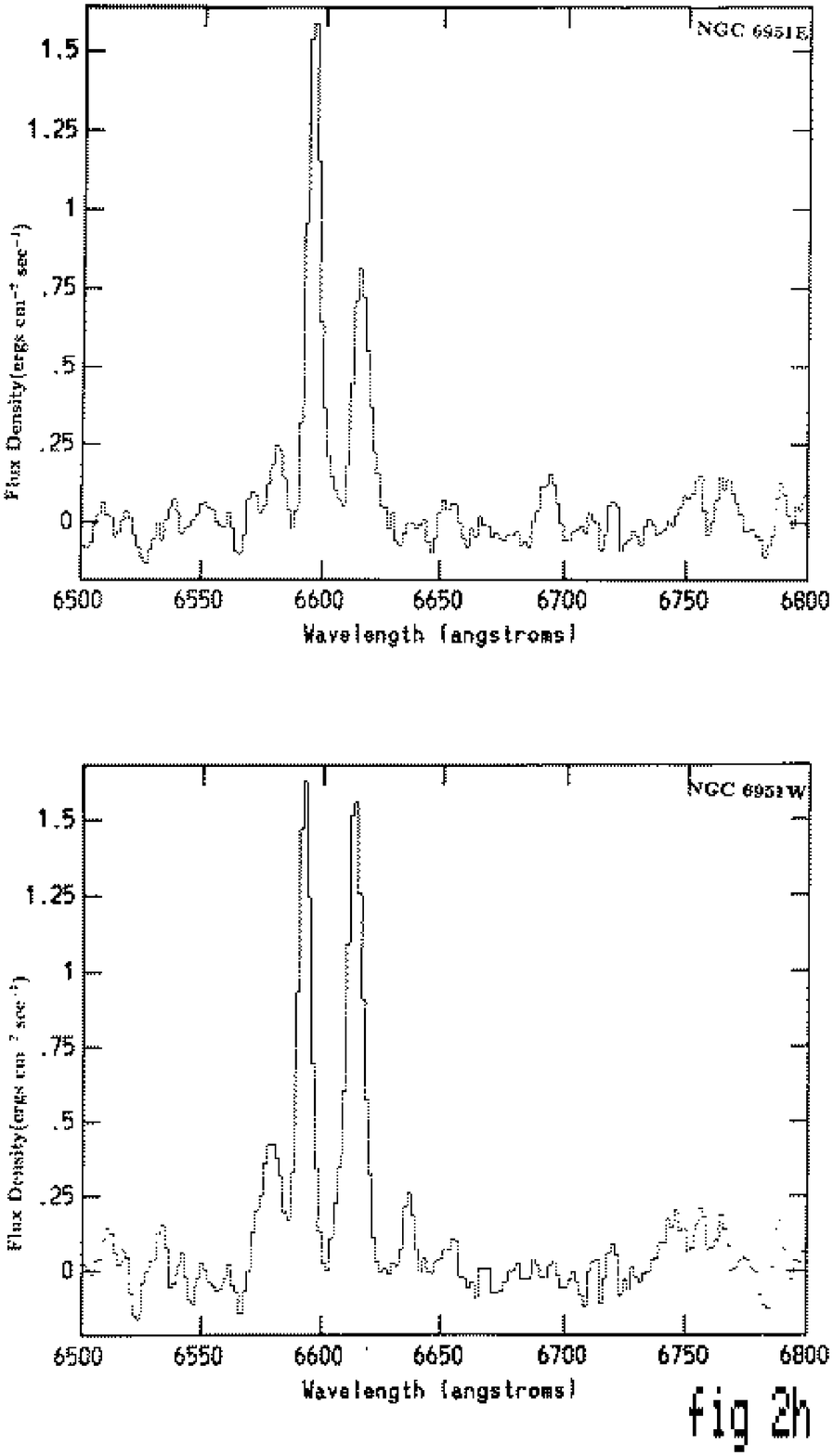}
\caption{As in Fig. 2, but for NGC 6951.}
\label{fig2h}
\end{figure}

% table 4
\begin{deluxetable}{lrrrrr}
%\footnotesize
\scriptsize
\tablecaption{East-West Velocity differences and line ratios.\label{tbl-4}}
\tablewidth{0pt}
\tablehead{
\colhead{Galaxy} & \colhead{${\Delta}V_{e-n}$\tablenotemark{a}}  
& \colhead{${\Delta}V_{e-w}$\tablenotemark{a}}   
& \colhead{${\Delta}V_{n-w}$\tablenotemark{a}} 
& \colhead{[SII]/[SII]\tablenotemark{b}} 
& \colhead{n$_e$~cm$^{-3}$\tablenotemark{c}}
}\startdata 
3504 & -39  & -39  & 0   &  1.0 & $\sim1000\pm600$ \nl
4314 &      & -121 &     &  0.6 & $\sim4500\pm1000$ \nl
4691 & -20  & -10  &  10 &  1.1 & $\sim750\pm250$  \nl
5135 &   7  & -19  & -26 &  1.3 & $\sim400\pm200$  \nl
5383 & 75   & 230  & 155 &  1.2 & $\sim500\pm200$  \nl
5534 &      & -22  &     &  1.4 & $<100$   \nl
5915 & 68   & 143  & 75  &  1.0 & $\sim1000\pm600$ \nl
6951 & 59   & 138  & 79  &  0.9 & $\sim1600\pm600$ \nl    
\enddata
\tablenotetext{a}{H$\alpha$ velocity differences: e-n means east minus nucleus; e-w means east minus west and n-w means nucleus minus west. All velocities are taken from tables 2, 3 and 4.}
\tablenotetext{b}{Sulfur line ratio [SII] $\lambda$  6716.4 \AA over 
[SII] $\lambda$ 6730.8\AA }
\tablenotetext{c}{Approximate electron densities (\cite{ost89})}
\end{deluxetable}

% table 5
\begin{deluxetable}{lrr}
%\footnotesize
\scriptsize
\tablecaption{Line ratios.\label{tbl-5}}
\tablewidth{0pt}
\tablehead{
\colhead{Galaxy} & \colhead{[SII]/[SII]\tablenotemark{a}} 
& \colhead{n$_e$~cm$^{-3}$\tablenotemark{b}}
}\startdata 
4123 &   1.1 & $\sim750\pm250$ \nl
4477 &   0.7 & $\sim3500\pm750$ \nl
5347 &   1.0 & $\sim1200\pm600$ \nl
5430 &   1.2 & $\sim500\pm200$ \nl
5597 &   1.5 & $\leq100$ \nl
5691 &   0.9 & $\sim1600\pm600$ \nl
5728 &   1.2 & $\sim500\pm200$   \nl
5757 &   1.0 & $\sim1200\pm600$  \nl
6239 &  $\sim2.0$ & $<100$ \nl
6907 &  $\sim1.3$ & $\sim300\pm200$ \nl
\enddata
\tablenotetext{a}{Sulfur line ratio [SII] $\lambda$ 6716.4 \AA over 
[SII] $\lambda$ 6730.8 \AA }
\tablenotetext{b}{Approximate electron densities (\cite{ost89})}
\end{deluxetable}

% table 6
\begin{deluxetable}{lrrcr}
\scriptsize
\tablecaption{Estimates of Dynamical Central Masses.\label{tbl-6}}
\tablewidth{0pt}
\tablehead{
\colhead{Galaxy} & \colhead{Distance\tablenotemark{a}}  
& \colhead{R(pc)}
& \colhead{$V=V_{obs}/\sin(i)$\,\tablenotemark{b}}
& \colhead{M$_{dyn}$\tablenotemark{c}} 
}\startdata 
3504 & 26.5 &  560  &  83 & $8.5~10^8$   \nl
4314 & 10.0 &  390  & 115 & $1~10^9$   \nl
4691 & 22.5 &  480  &  34 & $1.3~10^8$ \nl
5135 & 53.2 &  755  &  24 & $1~10^8$ \nl
5383 & 37.8 & 1740  & 185 & $1.4~10^{10}$ \nl
5534 & 35.0 & 1730  &  25 & $2.4~10^8$    \nl
5915 & 33.7 & 1430  & 108 & $4~10^9$    \nl
6951 & 24.1 &  595  & 170 & $4~10^9$    \nl    
\enddata
% Text for table footnotes follows the tabular data and must be inside the
% deluxetable environment.  Note that it is OK to put \ref's in 
% \tablenotetext's.
\tablenotetext{a}{in Mpc from Tully (1988).}
\tablenotetext{b}{True velocity difference, in km s$^{-1}$, between the nucleus and the 
west (east) side in (NGC 4691), NGC 5135, NGC 5383, NGC 5915 and NGC 6951. True velocity half difference between the east and west sides in NGC 4314 and in NGC 5534, for which we assumed an inclination of 30$^{\circ}$.}
\tablenotetext{c}{Dynamical mass of each galaxy inside a radius R, in units of solar masses: $M_{dyn} = 233.7~R(pc)~V^2(km s^{-1})^2$}
\end{deluxetable}

Not all the galaxies had emission away from the central pixels which correspond 
to the nucleus. Our original list of galaxies was 26, but we only detected 
emission lines in the spectra of 18 of them. Eight out of 18 showed emission 
from the east and west side of the central regions, and 6 out of 8 showed 
emission from their compact nuclear region as well. In the end we extracted 
three different final spectra from each of the eight galaxies. In order to 
extract the final spectra we used the IRAF task apall where we co-added the 
pixels that correspond to each region. The numbers of pixels that were co-added 
to create each spectrum are given in the third column of Table 2. In none of 
the galaxies did we detect emission from the complete disk and thus no rotation 
curve could have been produced. Figures \ref{fig2a}, \ref{fig2b}, \ref{fig2c}, \ref{fig2d}, \ref{fig2e}, \ref{fig2f}, \ref{fig2g}, \ref{fig2h} shows the spectra of the eight galaxies from their eastern and western circumnuclear regions. Ten galaxies did not show any bright emission away  from their central pixels and a final spectrum for each of them was obtained after adding the central 5 to 20 pixels. 

The task ``splot''  was used to measure the strengths, central wavelengths, 
widths, and fluxes of the emission lines. The central wavelengths, widths and 
fluxes of emission lines were estimated by fitting gaussian curves with the 
IRAF task splot. For this purpose we fitted multiple curves in order to account 
for blended broad lines and when possible corroborated the values by fitting 
individual lines. The uncertainty for the velocity centroids of the emission 
lines amounts to about 25 km s$^{-1}$ as a result of wavelength calibration and
gaussian fitting. The instrumental spectral resolution was estimated by 
calculating the full width at half maximum of emission lines of a lamp at 
different wavelengths across the band. Our estimated value is 
$FWHM_{inst}\sim3.23\pm0.1$ \AA $\,$  or about 145$\pm5$ km s$^{-1}$. The 
tabulated line widths were corrected for instrumental resolution by subtracting 
the instrumental width in quadratures from the observed width: 
FWHM$^2_{real}$=FWHM$^2_{obs}$--~FWHM$^2_{inst}$.

\section{Results and Discussion}

In this section we present the results of our observations: first, we provide 
the estimates of central masses calculated from the observed velocities,
assuming that the ionized gas rotates on circular orbits around the compact nucleus and lies on the plane of each galaxy's disk. The velocities may also be 
useful in deriving the properties of resonances using density wave theory and 
theory of orbits in order to associate circumnuclear structures with Inner 
Lindblad Resonances (see Binney \& Tremaine 1987 and Contopoulos \& Grosb{\o}l 
1989). The spatial distribution of H${\alpha}$+[NII] of these barred galaxies, except for NGC 5383, has been reported by \cite{pog89a,pog89b} and 
\cite{gar96}. Then, the line ratios and electron densities can be derived.

\subsection{Velocities and Inner Masses}

Table 2 lists the heliocentric velocities, in km s$^{-1}$, for [NII]${\lambda 
6548.1}$, H$\alpha$ ${\lambda 6562.8}$, [NII]${\lambda 6583.4}$ and 
[SII]${\lambda 6716.4}$ and ${\lambda 6730.8}$ lines from the 8 galaxies we 
where were able to isolate the emission from the eastern and western sides. 
Table 3 lists the heliocentric velocities for other 10 galaxies with no 
emission away from the central pixels in our slit.

Our observations indicate that the different systemic velocities reported in 
the literature for NGC 4314 ($v_{sys}=883$ kms$^{-1}$ [Sandage \& Tammann 
1987]; $v_{sys}=1004$ kms$^{-1}$ [Smith et al. 1987; Strauss et al. 1992]) 
are most likely the result of their position of the slit relative to the 
circumnuclear structure. The velocities agree with the observed velocities for 
the eastern and western regions of the circumnuclear structure (see Table 2).
Our optical velocities are somewhat smaller then the velocities 
obtained from CO lines. In particular, the eastern CO emission peaks at around  
$v_{east}\approx920$ km s$^{-1}$ and the western CO emission peaks at around 
$v_{west}\approx1030$ km s$^{-1}$ (\cite{ben96}). The differences between 
optical and CO lines may be the result of hydrodynamic processes at play, e.g. 
molecular outflows, supernovae events, etc., that may produce a different 
velocity for the molecular and atomic components.

In NGC 5430, we only detected a velocity of $v\approx 3065$ km s$^{-1}$, which 
is quite different from the reported systemic optical velocity of 
$v_{sys}=2875$ km s$^{-1}$. On the other hand, the CO lines have $v_{NE}\approx2893$ km s$^{-1}$ and $v_{SW}\approx3069$ km s$^{-1}$ 
(\cite{con97}). This leads us to conclude that our spectrum only showed the 
emission from the southwestern (SW) blob, and that the reported systemic
velocity is probably better associated to the NE blob.

Our velocities for NGC 4691 and NGC 6951 are in agreement with the velocities 
reported by Wiklind, Henkel \& Sage (1993) and by Boer \& Schulz (1993). The 
optical velocities for NGC 3504 differ by $\approx 100$ km s$^{-1}$ from the CO 
velocities. This discrepancy is likely the result of positioning the slit off 
the nucleus (e.g. \cite{ken92}). Table 4 lists the velocity differences for the
galaxies observed. 

We estimate the dynamical mass interior to the rings of ionized matter. 
Assuming that most of the mass interior to these rings is homogeneously 
distributed inside a radius R, the dynamical mass may be estimated from 
M$_{dyn}$(M$_\odot$)=233.7 R(pc) V$^2_{true}$ (km s$^{-1}$)$^2$, where V$_{true}=$ 
V$_{obs}/\sin (i)$, with V$_{obs}$ the observed velocity and $i$ is the 
inclination angle of the circumnuclear structure with respect to the plane of 
the sky. We are assuming, not necessarily being correct, that the circumnuclear 
structure is on the plane of a given galaxy and its inclination is the same as 
the inclination of the galaxy disk. This assumption is based primarily on the 
idea that circumnuclear structures are at distances of about 300 pc up to 1 kpc 
from the compact nucleus and such a large structure would be unstable if being 
inclined with respect to the plane of the disk. In addition, we are implicitly assuming circular orbits 
for the gas in the rings, which might not be true especially in the presence of 
a pronounced bar.

Table 6 gives the estimated dynamical masses, M$_{dyn}$, assuming as we said 
before that the ionized gas in circumnuclear regions lies in the plane of the 
galaxy's disk and that they follow circular orbits as a response to a central 
axisymmetric gravitational potential. As mentioned earlier if circumnuclear structures (ie. rings) are to be explained as Inner Lindblad Resonances, more observations are needed in order to have the complete radial velocity field (rotation curves) and to compute epicyclic frequencies, $\kappa(r)$. We estimate an error of about a factor of two on the masses presented in Table 6.

\subsection{Line Ratios and Electron Densities}

\subsubsection{[NII]/H$\alpha$ ratio}

Burbidge \& Burbidge (1962) showed that the regions in external galaxies with 
H$\alpha$ and [NII] emission lines, and a line intensity ratio of 
[NII]$\lambda~$6583/H$\alpha\sim0.33$, have a degree of excitation that is 
similar to the extended HII regions in our own galaxy. They also pointed out 
that the observed ratio is larger than 1 in the nucleus of some galaxies. More 
recent observations of several emission lines,  with better sensitivity and 
spectral coverage, have been used to determine the excitation mechanisms in 
different objects (\cite{bal81,ost89}). In particular, the emission lines used 
are: [NeV]$\lambda~3426$ \AA, [OII]$\lambda~3727$ \AA, H$\beta~\lambda~4861$ 
\AA, HeII $\lambda~4686$ \AA, [OIII]$\lambda~5007$ \AA, [OI] $\lambda~6300$ 
\AA, H$\alpha\lambda~6563$ \AA, [NII]$\lambda~6584$ \AA. Depending on the line 
ratios, the excitation mechanisms for the line-emitting gas could be: a) 
photoionization by OB stars, b) photoionization by a power-law continuum 
source, c) collisional ionization from shock waves, and d) a combination of the 
above. 

Actually, the nuclear regions are usually excited by a power-law source but the
circumnuclear rings are excited by UV radiation from recently formed stars. 
Recent spectroscopic surveys of spiral galaxies show low-ionization emission, 
suggesting that a flat-spectrum (power law) radiation field photoionizes the 
nuclear gas (Keel 1983a,b,c;\cite{ken83,fil85}; Ho, Filippenko \& Sargent 1995, 1997a,b; \cite{hoe97}). 
In the nucleus of Seyfert galaxies the same mechanism seems to operate and 
high-ionization emission iron lines are found from the central regions, such 
as: [Fe VII]$\lambda$5721 \AA , [Fe VII]$\lambda$6087 \AA , [Fe X]$\lambda$6374 
\AA \ and [Fe XIV]$\lambda$5303 \AA $\;$  (\cite{app88}). In contrast, several 
observations support the idea that circumnuclear rings are regions of massive 
OB star formation: e.g., those  with bright H$\alpha$ emission, 10$\mu$ 
emission, radio continuum emission, and CO mm line emission (e.g. 
\cite{hum87,gar91a,gar91b,ken92,tel93,vil95,ben96,con97}). This is in line with
the expectations from galactic dynamics, where large accumulations of material 
can occur at regions near Lindblad resonances, hence, providing suitable 
conditions for triggering the formation of stars. Our results corroborate these
views, and we find that the photionizing agents for the nuclear gas and the circumnuclear rings are of different nature.

The line ratios [NII]/H$\alpha$ and [SII]/H$\alpha$ were computed separately 
for the eastern and western regions for all galaxies with circumnuclear 
structures.  The [NII]$\lambda~$6548/H$\alpha$ ratio is on average $\sim 0.12$ 
in the eastern side, $\sim0.16$  in the western side (Table 2). The ratio is 
$\sim0.25$ for galaxies without extranuclear emission (Table 3).  The 
[NII]$\lambda~$6583/H$\alpha$ ratio is on the average $\sim0.43$ in the eastern 
side and $\sim0.58$ in the western side. The average ratio is $\sim0.78$ for 
galaxies presenting spectra from their central regions without circumnuclear 
structures. The values found are in agreement with the values found in other 
barred galaxies like NGC 1097 (\cite{phi84}).

\begin{figure}[!t]
\epsfxsize=8cm
\epsffile{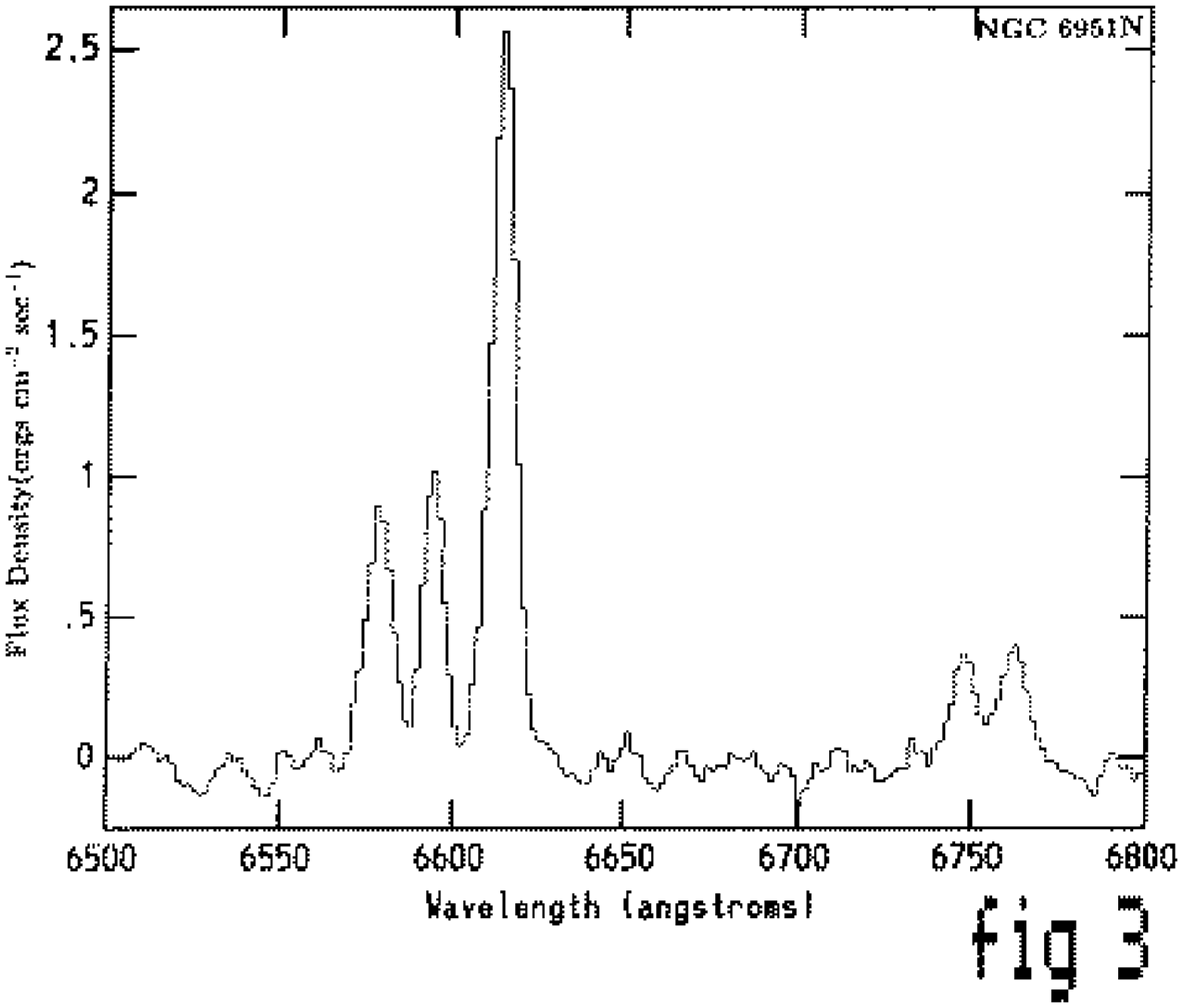}
\caption{ Spectrum from the compact nucleus in the barred galaxy NGC 6951 
calibrated in wavelength, sky, flux and redshift. Notice that the 
H$\alpha$ intensity is much lower than the [NII]$\lambda$6583. Flux units are $10^{-15}$~ergs s$^{-1}$~cm$^{-2}$~\AA$^{-1}$.}
\label{fig3}
\end{figure}

In the case of NGC 4314 and NGC 6951, with clear circumnuclear rings, the line 
ratio was found different from each side of the ring. The 
[NII]$\lambda~$6583/H$\alpha$ ratios in NGC 4314 are 0.45 and 0.85 while in 
NGC 6951 they are 0.48 and 0.93 from their eastern and western circumnuclear 
regions, respectively. These values suggest different intrinsic physical 
conditions as a result of local star formation and evolution in each side of 
the ring. We believe that the ratio of [NII]$\lambda6583$/H$\alpha$ of 1.01 
quoted by \cite{ho97b} for NGC 4314 represents the ratio from the western side 
of the circumnuclear ring and not necessarily from the compact nucleus since
images suggest that there is no H$\alpha$ emission from the compact nucleus 
(\cite{pog89a,gar96}). Additionally, the ratio [NII]$\lambda6583$/H$\alpha$ 
from the compact nucleus of NGC 6951 is about 2, in agreement with previous 
values reported (\cite{ho97b}). Figure~\ref{fig3} shows the spectrum corresponding to the compact nucleus of NGC 6951. NGC 4477 and NGC 5728 also show a ratio of  [NII]$\lambda~$6583/H$\alpha$ of $\sim2.4$ and $\sim1.4$ respectively, in  accord to values found previously (\cite{phi84,sch88,ho97b}). These galaxies  are classified as Seyfert 2 galaxies with physical characteristics that account 
for the ratio observed in the nucleus (\cite{ken89,ho97a,ho97b}).

\subsubsection{[SII]/H$\alpha$ and [SII]$\lambda6716$/[SII]$\lambda6731$}

The ratio of [SII]/H$\alpha$ is found in the range from 0.07 to 0.2 with no 
obvious difference in either side of circumnuclear regions. It is in the range 
from 0.1  to 0.3 from the compact nucleus in galaxies with circumnuclear 
structures and it is 0.1 to 0.4 in central regions of galaxies without 
circumnuclear regions (see Tables 2 and 3). The lower values seem to be from 
the eastern and western side of the nuclei, but further observations with 
longer integration times and wider slits are needed to confirm this. Ratios 
with larger values from nuclei and low values from circumnuclear regions have 
been found in NGC 1097 (\cite{phi84}) and NGC 5728 (\cite{sch88}).

The [SII]$\lambda6716$/[SII]$\lambda6731$ ratio was found to be in the interval
from 0.6 to 1.4, suggesting electron densities of a few hundred particles per 
cm$^{3}$ (see Tables 4 and 5; e.g. \cite{ost89}).

\section{Summary}

We have presented new long slit spectroscopic observations with [NII], 
H$\alpha$ and [SII] emission lines from 18 RSA  barred spiral 
galaxies, eight of which have structures around the nucleus. The main results
are the following: a) We were able to determine the heliocentric optical 
velocities for the ionized gas from the compact nucleus, and the eastern and
western regions, in eight galaxies. With these velocities we have estimated the 
inner dynamical mass of each galaxy assuming that the gas move in circular
orbits on the plane of the galaxy. b) We were able to estimate the line ratios 
[NII]$\lambda6548$/H$\alpha$, [NII]$\lambda6583$/H$\alpha$, 
[SII]$\lambda6716$/H$\alpha$ and [SII]$\lambda6730$/H$\alpha$ for the eastern, 
western, and compact nuclear regions, separately, of eight galaxies. In 
particular, we found that in galaxies with clear circumnuclear rings, as in 
NGC 4314 and NGC 6951, the ratio [NII]$\lambda6583$/H$\alpha$ is lower in the 
eastern region than in the western region of the ring suggesting different 
local physical conditions as a result of star formation and evolution. c) In 
NGC 4477, NGC 5728 and NGC 6951 we found that the ratio 
[NII]$\lambda6583$/H$\alpha$ is larger than unity in the central nuclear 
regions, in agreement with values reported.
	
\section*{Acknowledgments}

We acknowledge useful comments from the referee that helped us to improve the 
final version of this paper. J.A.G-B acknowledges partial financial support 
from DGAPA (UNAM) and CONACYT  (971010), M\'exico, that allowed him to spend 
his sabbatical year at the Astronomy Department of the University of Minnesota 
and he would like to thank the Astronomy Department of the University of 
Minnesota for their hospitality where part of the original article was written. 
H.A. thanks the Spanish Ministry of Foreign Affairs for financial support. J.F. 
acknowledges the support given to this project by DGAPA-UNAM grant, CONACyT 
grants 400354-5-4843E and 400354-5-0639PE, and a R\&D Cray research grant. This 
research has made use of the NASA/IPAC extragalactic database (NED) which is 
operated by the Jet Propulsion Laboratory, Caltech, under contract with the 
National Aeronautics and Space Administration.

%======={\center \section*{References}}
%=======\setlength{\parindent}{-1.0\parindent}

%=======\
\onecolumn

\end{document}